\documentclass[a4]{article}
\usepackage{amsmath}
\usepackage{epsfig}
\allowdisplaybreaks

\usepackage{thophys}
\usepackage{thohacks}

\begin{document}
\renewcommand{\thefootnote}{\fnsymbol{footnote}}
\thispagestyle{empty}
\begin{titlepage}
\begin{flushright}
hep-ph/9805503 \\
TTP 98--24 \\
\today
\end{flushright}

\vspace{0.3cm}
\boldmath
\begin{center}
\Large\bf  Spectator Effects in Heavy Quark Effective Theory at
           $\mathcal O(1/m_Q^3)$
\end{center}
\unboldmath
\vspace{0.8cm}

\begin{center}
{\large Christopher Balzereit\footnote[5]{email:
    chb@particle.physik.uni-karlsruhe.de}}\\
{\sl Institut f\"{u}r Theoretische Teilchenphysik,
     Universit\"at Karlsruhe,\\ D -- 76128 Karlsruhe, Germany} 
\end{center}

\vspace{\fill}

\begin{abstract}
\noindent
We complete the one loop renormalization of the HQET lagrangian
at $\mathcal O(1/m_{Q}^{3})$ including four fermion operators
with two heavy and two light quark fields in the operator basis.
It is shown that as a consequence the short distance coefficients 
of the operators 
bilinear in the heavy quark field receive nontrivial corrections.
\end{abstract}
\end{titlepage}

\renewcommand{\thefootnote}{\arabic{footnote}}
\newpage

\section{Introduction}
\label{sec:introduction}
Heavy Quark Effective Field Theory
(HQET)~\cite{HQET_classic} has become a well
established theoretical tool for the description of
hadrons containing one heavy
quark~\cite{HQET_review}.  This derives from the fact that it is a
systematic expansion in inverse powers of the heavy quark mass $m_{Q}$ with
well defined and calculable coefficients.  Furthermore, its
realization of the spin and flavor symmetry of the low energy theory
is a phenomenologically powerful tool. 
The $1/m_Q$ expansion has already been applied sucessfully to 
phenomenological problems such as the determination of $V_{cb}$. 

Besides its phenomenological application HQET posseses 
interesting theoretical features which already show up if one studies the HQET 
lagrangian itself.  
On one hand it is relevant for power corrections
to the $1/m_{Q}$ expansion of hadron masses and on the other hand allows for 
a study of reparametrization invariance \cite{repar}. 

However, the complete construction of HQET 
has to include  radiative corrections. These are computable order by order 
in perturbation theory by matching HQET to full QCD at a 
perturbative scale $\mu = m_{Q}$
followed by the renormalization group running of the
respective operators. As indespensable ingredients 
both the coefficients at the matching scale and the 
anomalous dimensions of the operators are needed.

In this paper we concentrate on the one loop renormalization
of the lagrangian at $\mathcal O(1/m_{Q}^{3})$.
Results for the terms of $\mathcal O(1/m_{Q})$ and $\mathcal O(1/m_{Q}^{2})$
as well as a matching calculation for the terms of $\mathcal O(1/m_{Q}^{3})$
are already known \cite{eichten,grozin,balzi,blok,bauer,finke,manohar} .

In \cite{balzereit} the one loop anomalous dimensions of 
the effective lagrangian
at $\mathcal O(1/m_{Q}^{3})$ have been calculated. 
However, these results were incomplete in the sense that only
operators bilinear in the heavy quark field 
had been included in the operator basis.
In the present paper we complete the renormalization of 
the effective lagrangian 
extending the operator basis to four fermion operators composed of two
heavy and two light fields. 
At $\mathcal O(1/m_{Q}^{2})$ it is well known that the running
of the short--distance coefficient of the Darwin operator
is modified in the presence of such operators.
At higher orders this effect should be expected to cause 
more drastical consequences because of the large operator bases.
In fact, it will be shown that at $\mathcal O(1/m^{3}_{Q})$  
the coefficients of the bilinear operators receive
nontrivial corrections in the presence of heavy--light operators.   

A possible phenomenological application of our results is
the estimate of $\mathcal O(1/m_{Q}^{3})$--corrections
in the operator product expansion of the B--meson decay width. 
It is well known that to this order the width is 
significantly affected by spectator quark effects induced by 
heavy--light operators appearing in the OPE.
However, there are  also time ordered products 
of the effective lagrangian 
with the local operators in the OPE. 
At this stage heavy--light operators 
in the lagrangian and their short distance corrections which 
are the subject of this paper induce additional spectator effects.
 
This note is organized as follows: in section~\ref{sec:basis} we introduce
our operator basis and discuss its reduction
to a set of linearly independent operators in section~\ref{sec:EOM}.  
Our result for the anomalous dimensions 
is presented
in section~\ref{sec:anodim}.
Finally, we
present the logarithmic contributions to the short distance coefficients 
of the effective lagrangian
in section~\ref{sec:RG-flow} and
conclusions in section~\ref{sec:conclusions}.

\section{Operator basis}
\label{sec:basis}
The lagrangian of HQET can be written as 
\begin{equation}
\mathcal L = \bar h_v (ivD) h_v + \sum_{n = 1}^{\infty}\frac{1}{(2m_{Q})^n}
\mathcal L^{(n)} + \mathcal L_{\mathrm{light}}\,,
\end{equation}
where 
\begin{equation}
\mathcal L_{\mathrm{light}} = -\frac{1}{4}F^{a}_{\mu\nu}F^{a\mu\nu}
 + \bar q (i\fmslash{D} - m_{q}) q
\end{equation}
describes the dynamics of the light degrees of freedom.
For simplicity we consider only one  light quark with mass $m_{q}$.
Additional light flavor are taken care of by summing over
the index $q$. 

To lowest order in the heavy mass expansion
only one operator, $\bar h_v (ivD) h_v$, shows up
which has the celebrated spin-- and flavor--symmety properties.

However there are power corrections  
of $\mathcal O(1/m_{Q})$ to the 
heavy--quark limit which break these symmetries. They are given by 
a sum of operators of appropriate canonical dimension
multiplied by short distance coefficients:
\begin{equation}
\mathcal L^{(n)} = \sum_i C^{(n)}_i(\mu) \mathcal O^{(n)}_i(\mu)
\end{equation}
At given order $\mathcal O(1/m_Q^{n})$ all operators of the respective 
canonical dimension allowed by the symmetries of the effective
theory may contribute. 

First of all there are the operators bilinear in 
the heavy quark fields already appearing in the tree level lagrangian
(denoted H--operators).

Operators containing two heavy and an even number of additional light quark
fields appear if one matches QCD amplitudes with heavy and light external
fields onto HQET to one loop.
Their inclusion   
accounts for spectator quark effects inside hadronic states,
between which these operators are sandwiched when calculating
physical matrix elements. However, since their massdimension
must be larger or equal to 6 these operators 
show up at $\mathcal O(1/m_{Q}^2)$ for the first time.
To this order 
they can be constructed by two heavy and two light quark fields.
At $\mathcal O(1/m_{Q}^3)$ a covariant derivative or a factor 
$m_q$ increases the mass dimension. In what follows these
operators are referred to as HL--operators.

Since we remain in the one particle sector of HQET operators
with four or more heavy quark fields are ommited.

To complete the operator basis we should also consider 
operators constructed of four light quark fields and the 
corresponding penguin operator. However, to one loop order 
these operators only mix with the HL--operators. Since we are focusing on the 
corrections to the  H--operators we therefore neglect such operators.

Before we proceed to $\mathcal O(1/m_{Q}^3)$
let us recall the operators appearing at lower orders.

At $\mathcal O(1/m_Q)$ we choose the conventional basis
\begin{equation}
\begin{aligned}
  \mathcal{O}^{(1)}_1
    &= \bar h_v (iD)^2 h_v \qquad
  \mathcal{O}^{(1)}_2
    &= \frac{g}{2} \bar h_v \sigma^{\mu \lambda} 
       F_{\mu \lambda} h_v \\
  \mathcal{O}^{(1)}_3
    &= \bar h_v (ivD)^2 h_v \,,
\end{aligned}
\end{equation}
and at $\mathcal O(1/m_Q^2)$
\begin{equation}
\begin{aligned}
  \mathcal{O}^{(2)}_1
    &= \bar h_v iD_{\mu} (ivD) iD^{\mu} h_v &\qquad
  \mathcal{O}^{(2)}_2
    &= \bar h_v i \sigma^{\mu \lambda} iD_{\mu}
       (ivD) iD_{\lambda} h_v  \\
  \mathcal{O}^{(2)}_3
    &= \bar h_v (ivD) (iD)^2 h_v &\qquad
  \mathcal{O}^{(2)}_4
    &= \bar h_v (iD)^2 (ivD) h_v  \\
  \mathcal{O}^{(2)}_5
    &= \bar h_v (ivD)^3 h_v &\qquad
  \mathcal{O}^{(2)}_6
    &= \bar h_v (ivD) i \sigma^{\mu \lambda} iD_{\mu}iD_{\lambda}
       h_v  \\
  \mathcal{O}^{(2)}_7
    &= \bar h_v i \sigma^{\mu \lambda}
       iD_{\mu} iD_{\lambda} (ivD) h_v \,. & 
\end{aligned}
\end{equation}
The definition of the covariant derivative 
$iD = i\partial + g_sT^{a}A^{a}$ and field strength tensor
$F_{\mu\nu}^{a}T^{a} = -i/g_{s} [iD_{\mu},iD_{\nu}]$ follows usual conventions.

We choose the HL--operators at $\mathcal O(1/m_{Q}^2)$
as
\begin{eqnarray}
\mathcal M^{(2)s/o}_{1} \! &=& \! g_{s}^{2}[\bar q \mathcal C^a_{s/o}q] 
                       [\bar h_v \mathcal C^a_{s/o} h_v]\nonumber \\
\mathcal M^{(2)s/o}_{2} \! &=& \! g_s^2 [\bar q\fmslash{v} \mathcal C^a_{s/o}q] 
                            [\bar h_v \mathcal C^a_{s/o} h_v]\nonumber  \\
\mathcal M^{(2)s/o}_{3} \! &=& \! g_s^2 [\bar q\gamma_{5} \gamma^{\mu}\mathcal C^a_{s/o}q] 
                            [\bar h_v \gamma_{5}\gamma_{\mu} \mathcal C^a_{s/o} h_v] \\
\mathcal M^{(2)s/o}_{4} \! &=&\! g_s^2 [\bar q i\sigma^{\mu\nu} \mathcal C^a_{s/o}q] 
                            [\bar h_v i\sigma_{\mu\nu}\mathcal C^a_{s/o} h_v]
\nonumber  
\end{eqnarray}
where $\mathcal C^a_{s} = 1$ in the color singlett and 
$\mathcal C^a_{o} = T^a$ in the color octett case.

At  $\mathcal O(1/m_Q^3)$ we find 13 local H--operators
contributing to physical matrixelements:
\begin{equation}
\begin{aligned}
\mathcal{O}^{(3)}_{1} &= \bar h iD_{\mu}(ivD)^2 iD^{\mu} h &\qquad
\mathcal O^{(3)}_{2} &= \bar h (iD)^2(iD)^2 h \\
\mathcal O^{(3)}_{3} &= \bar h iD_{\mu}(iD)^2 iD^{\mu}h &\qquad
\mathcal O^{(3)}_{4} &= \bar h iD_{\mu}iD_{\nu} iD^{\mu}iD^{\nu}h \\
\mathcal O^{(3)}_{5} &= \bar h i\sigma^{\mu\nu}iD_{\mu} (ivD)^2 iD_{\nu} h  &\qquad
\mathcal O^{(3)}_{6} &= \bar h i\sigma^{\mu\nu}iD_{\mu}iD_{\nu}(iD)^2 h \\
\mathcal O^{(3)}_{7} &= \bar h i\sigma^{\mu\nu}iD_{\rho}iD_{\mu}iD_{\nu}iD^{\rho}h &\qquad
\mathcal O^{(3)}_{8} &= \bar h i\sigma^{\mu\nu}(iD)^2 iD_{\mu}iD_{\nu}h \\
\mathcal O^{(3)}_{9} &= \bar h i\sigma^{\mu\nu}iD_{\mu}(iD)^2iD_{\nu}h &\qquad
\mathcal O^{(3)}_{10} &= \bar h i\sigma^{\mu\nu}iD_{\mu}iD_{\rho}iD_{\nu}iD^{\rho}h \\
\mathcal O^{(3)}_{11} &= \bar h i\sigma^{\mu\nu}iD_{\rho}iD_{\mu}iD^{\rho}iD_{\nu}h &\qquad
\mathcal O^{(3)}_{12} &=g_s^2 \bar h F^{a\mu\nu}F^a_{\mu\nu}  h \\
\mathcal O^{(3)}_{13} &=g_s^2 \bar h v_{\nu}v^{\rho}F^{a\mu\nu}F^a_{\mu\rho}h & 
\end{aligned}
\end{equation}
Note that, in contrast to the case of the lower order operators, we omit
operators
vanishing by the heavy quark equation of motion (EOM) 
\begin{equation}
(ivD)h_v =0
\end{equation}
from the very beginning.
This is justified since we are not going to insert such operators in
time ordered products of order $\mathcal O(1/m_Q^{4})$ or higher.
We may not remove such operators
from the operatorbases at $\mathcal O(1/m_{Q})$ and $\mathcal O(1/m_{Q}^{2})$
since time ordered products of these operators at 
$\mathcal O(1/m_{Q}^{3})$ may be identical to local physical operators
and contribute to their anomalous dimensions (see section \ref{sec:CI}).
However, the following operators are needed as parts of the gluon EOM
and though being irrelevant we will keep them in the basis: 
\begin{align}
\mathcal O^{(3)}_{14} &= \bar h_{v}(ivD)^{2}(iD)^{2} h_{v} & \mathcal
O^{(3)}_{15} &= \bar h_{v} (ivD)(iD)^{2}(ivD)
h_{v} \nonumber\\
\mathcal O^{(3)}_{16} &= \bar h_{v}(iD)^{2}(ivD)^{2} h_{v} & \mathcal
O^{(3)}_{17} &= \bar h_{v} iD_{\mu}(ivD)iD^{\mu}(ivD)
h_{v} \nonumber\\
\mathcal O^{(3)}_{18} &= \bar h_{v}(ivD)iD_{\mu}(ivD)iD^{\mu} h_{v} & &
\nonumber
\end{align}

To construct a basis of HL--operators at $\mathcal O(1/m_Q^3)$ 
we let a covariant derivative  
act either on a light or a heavy quark field.
A further classification has to be taken 
according to the direction in which the covariant derivative acts.
We define 
\begin{equation}
iD^{+}_{\mu} = i\overrightarrow{\partial}_{\mu} + g_s A_{\mu}^a T^a \,
\qquad 
iD^{-}_{\mu} = i\overleftarrow{\partial}_{\mu} - g_s A_{\mu}^a T^a
\end{equation}
and it is understood that in the octett case the covariant derivative stands
left/right of the color matrix when acting to the left/right.
With these definitions the HL--operators of
mass dimension 7 are choosen as:
\begin{eqnarray}
\mathcal M^{(3h)s/o}_{1\pm} \! &=&\! \!\pm g_s^2[\bar q \mathcal C^a_{s/o}q] 
                           [\bar h_v \mathcal C^a_{s/o} (ivD^{\pm}) h_v]   \nonumber \\
\mathcal M^{(3h)s/o}_{2\pm} &=&\pm g_s^2[\bar q \fmslash{v}\mathcal C^a_{s/o}q] 
                           [\bar h_v \mathcal C^a_{s/o} (ivD^{\pm}) h_v]  
\nonumber \\
\mathcal M^{(3h)s/o}_{3\pm} \! &=& \!\pm g_s^2[\bar q\gamma_{\mu} \mathcal C^a_{s/o}q] 
                           [\bar h_v i\sigma^{\mu\nu}\mathcal C^a_{s/o}
                           iD_{\nu}^{\pm} h_v]   \nonumber \\
\mathcal M^{(3h)s/o}_{4\pm} \! &=& \!\pm g_s^2[\bar q\gamma^{\mu} \mathcal C^a_{s/o}q] 
                           [\bar h_v \mathcal C^a_{s/o} iD_{\mu}^{\pm} h_v]   \nonumber \\
\mathcal M^{(3h)s/o}_{5\pm} \! &=& \!\pm g_s^2[\bar q i\sigma_{\mu\lambda}v^{\lambda}\mathcal C^a_{s/o}q] 
                           [\bar h_vi\sigma^{\mu\nu} \mathcal C^a_{s/o} iD_{\nu}^{\pm} h_v]   \nonumber \\
\mathcal M^{(3h)s/o}_{6\pm} \! &=& \!\pm g_s^2[\bar qi\sigma^{\mu\lambda}v_{\lambda} \mathcal C^a_{s/o}q] 
                           [\bar h_v \mathcal C^a_{s/o} iD_{\mu}^{\pm} h_v]   \nonumber \\
\mathcal M^{(3h)s/o}_{7\pm} \! &=& \!\pm g_s^2[\bar q \gamma_{5}\fmslash{v}\mathcal C^a_{s/o}q] 
                           [\bar h_v\gamma_{5} \mathcal C^a_{s/o} i\fmslash{D}^{\pm} h_v]   \nonumber \\
\mathcal M^{(3h)s/o}_{8\pm} \! &=& \!\pm g_s^2[\bar q\gamma_{5}\gamma^{\mu} \mathcal C^a_{s/o}q] 
                           [\bar h_v\gamma_{5}\gamma_{\mu}  \mathcal C^a_{s/o} (ivD^{\pm}) h_v]   \nonumber \\
\mathcal M^{(3h)s/o}_{9\pm} \! &=& \!\pm g_s^2[\bar q\gamma_{5} \mathcal C^a_{s/o}q] 
                           [\bar h_v\gamma_{5} \mathcal C^a_{s/o} i\fmslash{D}^{\pm} h_v]   \nonumber \\
\mathcal M^{(3h)s/o}_{10\pm} \! &=& \!\pm g_s^2[\bar qi\sigma^{\mu\nu} \mathcal C^a_{s/o}q] 
                           [\bar h_v i\sigma_{\mu\nu}\mathcal C^a_{s/o} (ivD^{\pm}) h_v]    \\
\mathcal M^{(3l)s/o}_{1\pm} \! &=& \!\pm g_s^2[\bar q \mathcal C^a_{s/o} i\fmslash{D}^{\pm} q] 
                           [\bar h_v \mathcal C^a_{s/o} h_v]   \nonumber \\
\mathcal M^{(3l)s/o}_{2\pm} \! &=& \!\pm g_s^2[\bar q \mathcal C^a_{s/o} (ivD^{\pm}) q] 
                           [\bar h_v \mathcal C^a_{s/o} h_v]   \nonumber \\
\mathcal M^{(3l)s/o}_{3\pm} \! &=& \!\pm g_s^2[\bar q \fmslash{v} \mathcal C^a_{s/o} (ivD^{\pm}) q] 
                           [\bar h_v \mathcal C^a_{s/o} h_v]   \nonumber \\
\mathcal M^{(3l)s/o}_{4\pm} \! &=& \!\pm g_s^2[\bar q i\sigma^{\lambda\nu}v_{\lambda}\mathcal C^a_{s/o} iD_{\nu}^{\pm} q] 
                           [\bar h_v \mathcal C^a_{s/o} h_v]   \nonumber \\
\mathcal M^{(3l)s/o}_{5\pm} \! &=& \!\pm g_s^2[\bar q i\sigma^{\mu\nu}\mathcal C^a_{s/o} (ivD^{\pm}) q] 
                           [\bar h_v i\sigma_{\mu\nu}\mathcal C^a_{s/o} h_v]   \nonumber \\
\mathcal M^{(3l)s/o}_{6\pm} \! &=& \!\pm g_s^2[\bar q \gamma_{5}\fmslash{v}\mathcal C^a_{s/o} iD_{\mu}^{\pm} q] 
                           [\bar h_v \gamma_{5}\gamma^{\mu}\mathcal C^a_{s/o} h_v]   \nonumber \\
\mathcal M^{(3l)s/o}_{7\pm} \! &=& \!\pm g_s^2[\bar q \gamma_{5}\gamma^{\mu}\mathcal C^a_{s/o} (ivD^{\pm}) q] 
                           [\bar h_v \gamma_{5}\gamma_{\mu}\mathcal C^a_{s/o} h_v]   \nonumber \\
\mathcal M^{(3l)s/o}_{8\pm} \! &=& \!\pm g_s^2[\bar q \gamma_{5}\mathcal C^a_{s/o} iD_{\mu}^{\pm} q] 
                           [\bar h_v \gamma_{5}\gamma^{\mu}\mathcal C^a_{s/o} h_v]   \nonumber \\
\mathcal M^{(3l)s/o}_{9\pm} \! &=& \!\pm g_s^2[\bar i\sigma_{\lambda\nu} q \mathcal C^a_{s/o} iD_{\mu}^{\pm} q] 
                           [\bar h_v i\sigma^{\mu\nu}\mathcal C^a_{s/o} h_v]   \nonumber \\ 
\mathcal M^{(3l)s/o}_{10\pm} \! &=& \!\pm g_s^2[\bar q\gamma_{\nu}  \mathcal C^a_{s/o} iD_{\mu}^{\pm} q] 
                           [\bar h_v i\sigma^{\mu\nu}\mathcal C^a_{s/o} h_v]
\nonumber
\end{eqnarray}

To complete the set of local operators we also 
consider operators of lower dimension multiplied by an appropriate 
power of the light quark mass:
\begin{equation}\label{eq:mops}
m_q \mathcal O^{(1)}_i,  
m_q^{2} \mathcal O^{(1)}_i,  
m_q \mathcal O^{(2)}_i,  
m_q \mathcal M^{(2)s/o}_{1}
\end{equation}
These operators are needed as counterterms 
during renormalization as long as the light quark
is kept massive. However the 
mass dependent HL--operator can be removed with help of  the light--quark
EOM. This is not the case for the mass dependent H--operators 
which are induced
by penguin type diagrams.
They cause corrections proportional to powers of $m_q/m_Q$ to 
the short distance coefficients of the respective operators.
Note that the operators $\mathcal M^{(3h)s/o}_{i}$, $i = 1,2,8,10$
vanish by the heavy quark EOM. Instead of disregarding these
operators from the very beginnig we will keep them in the basis
since they show up in several operator identities.

In addition to the local operators there are the
time ordered products of dimension 7 composed of
lower dimension operators:
\begin{align}
\mathcal T^{(12)}_{ij} &=
i\Tprod{\mathcal O^{(1)}_{i},\mathcal O^{(2)}_{j}}& i=1,2,3\quad j = 1,\ldots,7
\nonumber\\
\mathcal T^{(111)}_{ijk}&= -\mathcal S_{ijk}
\Tprod{\mathcal O^{(1)}_{i},\mathcal O^{(1)}_{j},\mathcal O^{(1)}_{k}}
& i,j,k = 1,2,3 \quad i\leq j \leq k\\
\mathcal T^{(12hl)s/o}_{ij} &= 
i\Tprod{\mathcal O^{(1)}_{i},\mathcal M^{(2)s/o}_{j}}
& i = 1,2,3 \quad j = 1,\ldots,4 \nonumber
\end{align}   
The symmetry factor $\mathcal S_{ijk}$ equals $1 ,1/2$ or
$1/6$, if no, two or all inserted operators are identical.
Again a mass dependent T--product
\begin{equation} 
m_{q}(1 - \frac{1}{2}\delta_{ij}) 
i\Tprod{\mathcal O^{(1)}_{i},\mathcal O^{(1)}_{j}}
\end{equation}
contributes. 
Although the short distance coefficients of time ordered
products are the products of the coefficents of their
operator components, local operators are 
required for their renormalization. This in turn 
modifies the running of the coefficents of the local operators.

\section{Reduction of the operator basis}
\label{sec:EOM}
The operator basis presented in the previous section
is overcomplete, i.e. there exist several interdependencies
between the operators. To arrive at a physical basis
of linearely independent operators 
all redundant operators have to be removed, otherwise
artificial gauge dependencies may show up in the 
corresponding short distance coefficients \cite{balzereit}.
\subsection{Momentum conservation}
\label{sec:mom}
Since all operators are integrated over their argument,
momentum conservation is manifest.

In case of the H--operators the only effect is that
the direction in which the covariant derivative acts is irrelevant. 

The partial derivative of a HL--operator
of mass dimension 6 vanishes. Performing the differentiation 
explicitly yields relations among the HL--operators
of dimension 7:
\begin{eqnarray}
0 &=& \partial_{\mu}[\bar q \Gamma^{\mu\nu} q][\bar h_{v}\Gamma_{\nu} h_{v}]
\nonumber\\
 &=& [\bar q iD^{-}_{\mu}\Gamma^{\mu\nu} q][\bar h_{v}\Gamma_{\nu} h_{v}]
   + [\bar q \Gamma^{\mu\nu} iD^{+}_{\mu}q][\bar h_{v}\Gamma_{\nu} h_{v}]\\
 & &+\,[\bar q \Gamma^{\mu\nu} q][\bar h_{v}iD^{-}_{\mu}\Gamma_{\nu} h_{v}]
+[\bar q \Gamma^{\mu\nu} q][\bar h_{v}\Gamma_{\nu}iD^{+}_{\mu} h_{v}]
 \nonumber\\
0 &=& \partial_{\mu}[\bar q \Gamma_{\nu} q][\bar h_{v}\Gamma^{\mu\nu}
h_{v}]\nonumber \\
 &=& [\bar q iD^{-}_{\mu} \Gamma_{\nu} q][\bar h_{v}\Gamma^{\mu\nu} h_{v}]
 +[\bar q \Gamma_{\nu} q][\bar h_{v}\Gamma^{\mu\nu}iD^{+}_{\mu} h_{v}]\\
& &+\,[\bar q \Gamma_{\nu} q][\bar h_{v}iD^{-}_{\mu}\Gamma^{\mu\nu} h_{v}]
+ [\bar q \Gamma_{\nu} q][\bar h_{v}\Gamma^{\mu\nu}iD^{+}_{\mu} h_{v}]
\nonumber
\end{eqnarray}

Choosing appropriate Dirac matrices $\Gamma_{\mu\nu}$ and
$\Gamma_{\mu}$ this allows us to remove the
following operators from the basis
(in favour of the combination on the r.h.s):
\begin{eqnarray}
\mathcal M^{(3l)s/o}_{1+} &=&  \mathcal M^{(3l)s/o}_{1-}
                           - \mathcal M^{(3h)s/o}_{4+}
                           + \mathcal M^{(3h)s/o}_{4-} \nonumber\\
\mathcal M^{(3l)s/o}_{3+} &=&  \mathcal M^{(3l)s/o}_{3-}
                           - \mathcal M^{(3h)s/o}_{2+}
                           + \mathcal M^{(3h)s/o}_{2-} \nonumber\\
\label{eq:mom}
\mathcal M^{(3l)s/o}_{5+} &=&  \mathcal M^{(3l)s/o}_{5-}
                           - \mathcal M^{(3h)s/o}_{10+}
                           + \mathcal M^{(3h)s/o}_{10-} \\
\mathcal M^{(3l)s/o}_{6+} &=&  \mathcal M^{(3l)s/o}_{6-}
                           - \mathcal M^{(3h)s/o}_{7+}
                           + \mathcal M^{(3h)s/o}_{7-} \nonumber\\
\mathcal M^{(3l)s/o}_{7+} &=&  \mathcal M^{(3l)s/o}_{7-}
                           - \mathcal M^{(3h)s/o}_{8+}
                           + \mathcal M^{(3h)s/o}_{8-} \nonumber\\
\mathcal M^{(3l)s/o}_{10-} &=& - \mathcal M^{(3l)s/o}_{10+}
                           + \mathcal M^{(3h)s/o}_{3+}
                           - \mathcal M^{(3h)s/o}_{3-} \nonumber\\
\end{eqnarray}

\subsection{Contraction Identities}
\label{sec:CI}
Time ordered products, in which  a  $(ivD)h_{v}$--term acts on
an internal line, are identical to a combination
of local operators:
\begin{equation}
i\Tprod{\bar h_v F(iD) (ivD)h_v, \bar h_v G(iD) h_v} = 
  - \bar h_v F(iD) G(iD) h_v + \ldots
\end{equation}
 If only H--operators are involved such {\it contraction identities} have 
been studied in \cite{balzereit},
where it has been shown that all time ordered products
of this type are redundant
and can be removed in favour of local H--operators.
This way the anomalous dimensions of the local H--operators
receive corrections that are crucial for the gauge independence
of their short distance coefficients.
\newpage
In the presence of HL--operators additional 
contraction identies arise:
\begin{eqnarray}
\label{eq:4fCI}
i\Tprod{[\bar h_{v}(ivD)^{2} h_{v}],g_{s}^{2}[\bar q \Gamma^{\mu} q][\bar
  h_{v} \Gamma^{\prime}_{\mu} h_{v}]} &=& 
-g_{s}^{2}[\bar q \Gamma^{\mu} q][\bar h_{v}(ivD^{-}) \Gamma^{\prime}_{\mu}
h_{v}]\nonumber \\
& & -g_{s}^{2}[\bar q \Gamma^{\mu} q][\bar h_{v} \Gamma^{\prime}_{\mu}(ivD^{+}) h_{v}]
\end{eqnarray}
Since the T--product is only related to HL--operators vanishing by
the heavy quark EOM the additional contraction identities
are irrelevant.

\subsection{Light Quark EOM}
\label{sec:LEOM}
Using the equation of motion for the light quark 
\begin{equation}
i\fmslash{D}^{+} q = m_{q}q \quad\quad -\bar q i\fmslash{D}^{-} = m_{q}\bar q
\end{equation}
we can remove the mass dependent HL--operators in (\ref{eq:mops})
from the 
operator basis:
\begin{eqnarray}
\label{eq:lightEOM}
m_{q}[\bar q \Gamma^{\mu}q][\bar h_{v} \Gamma^{\prime}_{\mu} h_{v}]
 &=& [\bar q \Gamma^{\mu}i\fmslash{D}^{+}q][\bar h_{v} \Gamma^{\prime}_{\mu} h_{v}]\\
 &=& -[\bar q i\fmslash{D}^{-} \Gamma^{\mu}q][\bar h_{v} \Gamma^{\prime}_{\mu}
 h_{v}] \nonumber
\end{eqnarray}
Taking a symmetrical combination of both equations (\ref{eq:lightEOM}) 
and choosing the dirac matrices $\Gamma^{\mu}$ and $\Gamma^{\prime}_{\mu}$ 
appropriately we 
derive the relations
\begin{eqnarray}
m_{q} \mathcal M^{(2)s/o}_{1} &=& -\frac{1}{2}\mathcal M^{(3h)s/o}_{4+}
                               +\mathcal M^{(3l)s/o}_{1-}
                              +\frac{1}{2} \mathcal M^{(3h)s/o}_{4-}
                               \nonumber \\
m_{q} \mathcal M^{(2)s/o}_{2} &=&\frac{1}{2} \mathcal M^{(3l)s/o}_{2+}
                              -\frac{1}{2} \mathcal M^{(3l)s/o}_{4+}
                              +\frac{1}{2} \mathcal M^{(3l)s/o}_{2-}
                              +\frac{1}{2} \mathcal M^{(3l)s/o}_{4-}
                               \nonumber \\ \label{eq:lightEOM1}
m_{q} \mathcal M^{(2)s/o}_{3} &=& \frac{1}{2}\mathcal M^{(3l)s/o}_{8+}
                              -\frac{1}{2} \mathcal M^{(3l)s/o}_{9+}
                              -\frac{1}{4} \mathcal M^{(3h)s/o}_{10+}
                             +\frac{1}{2} \mathcal M^{(3l)s/o}_{5-} \\
                           & &+\frac{1}{4} \mathcal M^{(3h)s/o}_{10-}
                              -\frac{1}{2} \mathcal M^{(3l)s/o}_{8-}
                              -\frac{1}{2} \mathcal M^{(3l)s/o}_{9-}
                               \nonumber \\
m_{q} \mathcal M^{(2)s/o}_{4} &=& \mathcal M^{(3h)s/o}_{7+}
                               -\mathcal M^{(3h)s/o}_{8+}
                              -2 \mathcal M^{(3l)s/o}_{6-}
                              +2 \mathcal M^{(3l)s/o}_{7-}
                               -\mathcal M^{(3h)s/o}_{7-}\nonumber \\
                               & &+\mathcal M^{(3h)s/o}_{8-}
                               +\mathcal M^{(3h)s/o}_{3+}
                               -\mathcal M^{(3h)s/o}_{3-}
                               \nonumber \,,
\end{eqnarray}
in which momentum conservation has been already used.

Subtracting both equations (\ref{eq:lightEOM}) yields relations
among the operators of dimension 7, written generically as:
\begin{equation}
0= [\bar q \Gamma^{\mu}i\fmslash{D}^{+}q][\bar h_{v} \Gamma^{\prime}_{\mu}
h_{v}] + [\bar q i\fmslash{D}^{-} \Gamma^{\mu}q][\bar h_{v} \Gamma^{\prime}_{\mu} h_{v}]
\end{equation}
Choosing appropriate dirac matrices $\Gamma_{\mu}$ and $\Gamma^{\prime}_{\mu}$
this identity allows us to remove the 
following operators from the basis (in favour of the operators on the r.h.s.):
\begin{eqnarray}
\label{eq:lightEOM2}
\mathcal M^{(3h)s/o}_{4+} &=&  \mathcal M^{(3h)s/o}_{4-} \nonumber \\
\mathcal M^{(3l)s/o}_{4+} &=&  -\mathcal M^{(3l)s/o}_{4-}
                                        +\mathcal M^{(3l)s/o}_{2+}
                                        -\mathcal M^{(3l)s/o}_{2-}
                                        \nonumber \\
\mathcal M^{(3h)s/o}_{10+} &=&  2\mathcal M^{(3l)s/o}_{8+}
                                        -2\mathcal M^{(3l)s/o}_{9+}
                                        +2\mathcal M^{(3l)s/o}_{8-}
                                        +2\mathcal M^{(3l)s/o}_{9-}
                                        +\mathcal M^{(3h)s/o}_{10-}
                                        \nonumber \\
\mathcal M^{(3l)s/o}_{10+} &=&  \frac{1}{2}\mathcal M^{(3h)s/o}_{3+}
                                    -\frac{1}{2}  \mathcal M^{(3h)s/o}_{7+}
                                    +\frac{1}{2}  \mathcal M^{(3h)s/o}_{8+}
                                    -\frac{1}{2}  \mathcal M^{(3h)s/o}_{3-}
                                    \nonumber \\
                                   & & +\frac{1}{2}  \mathcal M^{(3h)s/o}_{7-}
                                    +\frac{1}{2}  \mathcal M^{(3h)s/o}_{8-} 
\end{eqnarray} 

\subsection{Gluon EOM}
\label{sec:gluonEOM}
Taking the functional derivative of the lagrangian 
\begin{equation}
\mathcal L = -\frac{1}{4}F^{a}_{\mu\nu}F^{a\mu\nu}
 + \bar q (i\fmslash{D} - m_{q}) + \bar h_{v}(ivD) h_{v}
\end{equation}
with respect to the gluon field $A_{\lambda}^{b}$ 
yields the gluon EOM
\begin{equation}
\label{eq:gluonEOM1}
D_{\mu}^{ba} F^{a\mu\lambda} + g_{s}\bar q \gamma^{\lambda}T^{b}q 
                 + g_{s}v^{\lambda}\bar h_{v} T^{b} h_{v} =0\,,
\end{equation}
where $D^{ab} = \delta^{ab}\partial - g_s f^{abx}A^x$ is the 
covariant derivative in the adjoint representation.
We multiply $T^{b}v_{\lambda}$ 
and sandwich the resulting expression  between heavy quark spinors to get
\begin{equation}
\label{eq:gluonEOM2}
\bar h_{v} [iD_{\mu},[iD^{\mu},(ivD)]]h_{v} = 
   g_{s}^{2}[\bar  h_{v} T^{b} h_{v}][\bar q \fmslash{v} q] 
+ g_{s}^{2}[\bar h_{v}T^{b} h_{v}][\bar h_{v}T^{b} h_{v}]\,.
\end{equation}
The four--heavy--quark operator on the left hand side
can be omitted, since at least to one loop order 
such operators only mix with themselves
or appear as counterterms and therefore decouple from 
the renormalization group flow of the other operators.

In our conventions (\ref{eq:gluonEOM2})
reads
\begin{equation}
\label{eq:gluonEOM5}
\mathcal M^{(2)o}_{2} = -2\mathcal O^{(2)}_{1} + \mathcal O^{(2)}_{3}
                                +\mathcal O^{(2)}_{4}\,.
\end{equation}

To derive an analogous relation among operators of dimension 7
we multiply (\ref{eq:gluonEOM1}) with $T^{b}\Gamma^{\nu\lambda}$
and sandwich between $-\bar h_{v}iD^{-}_{\nu}$ and $h_{v}$
\begin{eqnarray}
\label{eq:gluonEOM3}
-g_{s}^{2}[\bar q \gamma_{\lambda}T^{b} q]
            [\bar h_{v}iD^{-}_{\nu}T^{b}\Gamma^{\nu\lambda} h_{v} ] &=&
-2 \bar h_{v} \Gamma^{\nu\lambda} iD_{\nu}iD_{\mu}iD_{\lambda}iD^{\mu} h_{v}
\nonumber \\
& &+ \bar h_{v} \Gamma^{\nu\lambda}iD_{\nu}iD_{\lambda}(iD)^{2}  h_{v}\\
& &+ \bar h_{v} \Gamma^{\nu\lambda}iD_{\nu}(iD)^{2} iD_{\lambda}
h_{v}\nonumber 
\end{eqnarray}
or between $\bar h_{v}$ and $iD^{+}_{\nu}h_{v}$
\begin{eqnarray}
\label{eq:gluonEOM4}
g_{s}^{2}[\bar q \gamma_{\lambda}T^{b} q] 
            [\bar h_{v}T^{b}\Gamma^{\nu\lambda}iD^{+}_{\nu} h_{v} ] &=&
-2 \bar h_{v} \Gamma^{\lambda\nu} iD_{\mu}iD_{\nu}iD^{\mu}iD_{\lambda} h_{v} 
\nonumber \\
& &+ \bar h_{v} \Gamma^{\lambda\nu}(iD)^{2} iD_{\nu}iD_{\lambda}  h_{v}\\
& &+ \bar h_{v} \Gamma^{\lambda\nu}iD_{\nu}(iD)^{2} iD_{\lambda}  h_{v}
\nonumber
\,.
\end{eqnarray}
Choosing 
$\Gamma_{\nu\lambda} = v_{\nu}v_{\lambda},g_{\nu\lambda},i\sigma_{\nu\lambda}$
yields the following relations:
\begin{eqnarray}
\mathcal M^{(3h)o}_{2-} &=& -2\mathcal O^{(3)}_{18} 
                        +\mathcal O^{(3)}_{14}
                        +\mathcal O^{(3)}_{15} \nonumber \\
\mathcal M^{(3h)o}_{2+} &=& -2\mathcal O^{(3)}_{17} 
                        +\mathcal O^{(3)}_{15}
                        +\mathcal O^{(3)}_{16} \nonumber \\ \label{eq:gEOMrel}
\mathcal M^{(3h)o}_{3-} &=& 2\mathcal O^{(3)}_{10} 
                        -\mathcal O^{(3)}_{6}
                        -\mathcal O^{(3)}_{9} \\
\mathcal M^{(3h)o}_{3+} &=& -2\mathcal O^{(3)}_{11} 
                        +\mathcal O^{(3)}_{8}
                        +\mathcal O^{(3)}_{9} \nonumber \\
\mathcal M^{(3h)o}_{4-} &=& -2\mathcal O^{(3)}_{4} 
                        +\mathcal O^{(3)}_{2}
                        +\mathcal O^{(3)}_{3} \nonumber \\ 
\mathcal M^{(3h)o}_{4+} &=& -2\mathcal O^{(3)}_{4} 
                        +\mathcal O^{(3)}_{2}
                        +\mathcal O^{(3)}_{3} \nonumber
\end{eqnarray}
Note that the relations involving $\mathcal M^{(3h)o}_{4\pm}$ are
consistent with the first relation in (\ref{eq:lightEOM2})
and the relations involving $\mathcal M^{(3h)o}_{2\pm}$
with the heavy quark EOM.

Taking the time ordered product of (\ref{eq:gluonEOM5}) with
a dimension 5 operator $\mathcal O^{(1)}_{i}$ yields
\begin{equation}
\label{eq:gluonEOM6}
\mathcal T^{(12hl)o}_{i2} = -2 \mathcal T^{(12)}_{i1}
 + \mathcal T^{(12)}_{i3}
 + \mathcal T^{(12)}_{i4}   \,,
\end{equation} 
where the last two T--products on the right hand side 
obey a contraction identity and are 
related to local operators $\mathcal O^{(3)}_i$. 
Note that in the case $i = 3$ (\ref{eq:gluonEOM6}) is consistent
with the contraction identity (\ref{eq:4fCI}), if 
the heavy quark EOM is applied.

\subsection{Reduced Operator basis}
\label{sec:redOp}
Applying the identities of the previous subsections 
to the full operator basis leaves us with 
the set of physical operators listed in figure \ref{fig:pos}.
\renewcommand{\arraystretch}{1.5}
In terms of this  set of linearely independent operators  
the effective lagrangian 
at $\mathcal O(1/m_Q^3)$ reads
\begin{equation}\label{eq:eff3}
\mathcal L^{(3)} = \vec C_{\mathcal B}\cdot \vec \mathcal B
                   +\vec C_{\mathcal O_m}\cdot \vec \mathcal O_m
                   +\vec C_{\mathcal M}\cdot \vec \mathcal M
                   +\vec C_{\mathcal T}\cdot \vec \mathcal T \,.
\end{equation}
The different types of operators and their coefficients are 
collected in vectors as defined in figure \ref{fig:pos}. 
\begin{figure}[ht]
\begin{center}
\begin{tabular}{|c|c|c|}\hline
operator& index & shorthand (number) \\
\hline 
$\mathcal O^{(3)}_{i}$ & $i=1,\ldots,13$ & \\\cline{1-2}
$\mathcal T^{(111)}_{ijk}$ & $i,j,k =1,2 \quad i \leq j \leq k$&$ \vec
\mathcal B \quad(21)$ \\\cline{1-2}
$\mathcal T^{(12)}_{ij}$ & $i,j =1,2 $ & \\\hline
$m^{2}_{q} \mathcal O^{(1)}_{i}$ & $i = 1,2$ & \\\cline{1-2}
$m_{q} \mathcal O^{(2)}_{i}$ & $i = 1,2$ &$\vec \mathcal O_{m}\quad(7)$ \\\cline{1-2}
$m_{q}\mathcal T^{(11)}_{ij}$ & $i,j =1,2 \quad i \leq j$ & \\ \hline
$\mathcal M^{(3h)s}_{i+}$ &$i=3,5,6,7,9$& \\\cline{1-2}
$\mathcal M^{(3h)s}_{i-}$ &$i=3,4,5,6,7,9$& \\\cline{1-2}
$\mathcal M^{(3h)o}_{i+}$ &$i=5,6,7,9$& \\\cline{1-2}
$\mathcal M^{(3h)o}_{i-}$ &$i=5,6,7,9$&$\vec \mathcal M \quad(43)$ \\\cline{1-2}
$\mathcal M^{(3l)s}_{i+}$ &$i=2,8,9$& \\\cline{1-2}
$\mathcal M^{(3l)s}_{i-}$ &$i=1,\ldots,9$& \\\cline{1-2}
$\mathcal M^{(3l)o}_{i+}$ &$i=2,8,9$& \\\cline{1-2}
$\mathcal M^{(3l)o}_{i-}$ &$i=1,\ldots,9$& \\\hline
$\mathcal T^{(12hl)s}_{ij} $ &$i = 1,2 \quad j = 1,\ldots,4$&  \\\cline{1-2}
$\mathcal T^{(12hl)o}_{ij} $ &$i = 1,2 \quad j = 1,3,4$& \raisebox{2.5ex}[0.5ex]{$\vec \mathcal T \quad(14)$} \\\hline
\end{tabular}
\end{center}
\caption{\label{fig:pos} Physical operator basis at $\mathcal O(1/m_Q^3)$.}
\end{figure}
\section{Anomalous dimensions}
\label{sec:anodim}
In order to compute the anomalous dimensions of the operators
$(\vec{\mathcal{B}},\vec{\mathcal{O}}_{m},
\vec{\mathcal{M}},\vec{\mathcal{T}})$, 
in general one has to calculate the 
pole parts of all divergent 1PI Greensfunctions with a certain operator
inserted and express the result in terms of tree--level operators
\begin{equation}
\label{eq:opexp}
\langle \mathcal A_{i} \rangle^{(1)}_{1PI}=
(\frac{\alpha}{\pi})\frac{1}{\epsilon}\sum_{j=0}\gamma_{ij}
     \langle \mathcal A_{j} \rangle^{(0)}_{1PI}
\end{equation}
where $\mathcal A_i$ runs over all operators of the basis.
From this one can directly read off the anomalous dimensions 
$\gamma_{ij}$.

\begin{figure}[ht]
\begin{center}
\begin{tabular}{|c|c|c|c|}
\hline
& diagram (example)&local counterterm&$\hat\gamma^{(3)}$ 
                                                             \\ \hline
                        &\mbox{
                    \leavevmode
                    \epsfxsize=3cm
                    \epsffile[180 640 410 730]{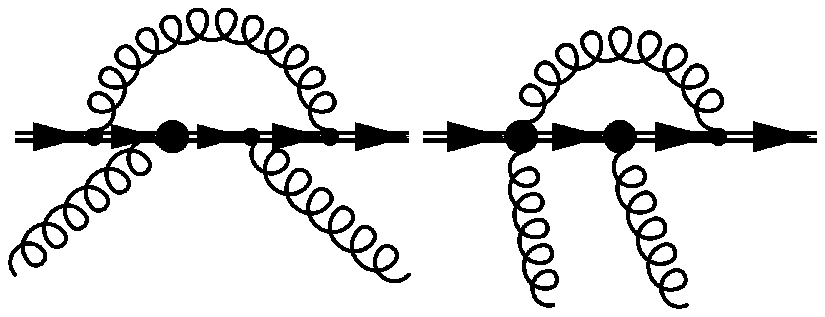}}
&\raisebox{2ex}[2ex]{$\mathcal{O}^{(3)}_{i}$}
  &\raisebox{2ex}[2ex]{$\hat{\gamma}_{\mathcal B\mathcal B}$}   \\ \cline{2-4}
\raisebox{2ex}[1ex]{$\vec \mathcal B $}      &
&$\mathcal M^{(3h/l)s/o}_{i\pm}$&
                        $\hat{\gamma}_{\mathcal B \mathcal M}$ \\ \cline{3-4}
                     &\raisebox{.5ex}[4ex]{
                    \leavevmode
                    \epsfxsize=3cm
                    \epsffile[200 670 410 750]{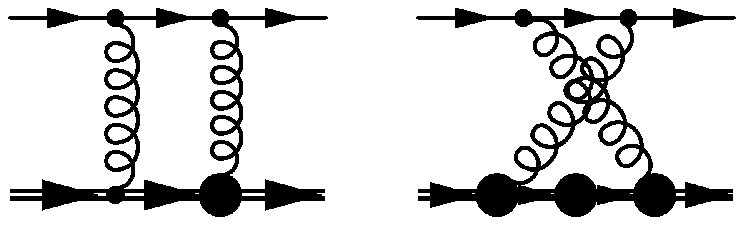}} 
                &$ \mathcal M^{(3h/l)s/o}_{i\pm} \rightarrow
                              \mathcal O^{(3)}_{i}$(gluon EOM)&
            $\hat{\gamma}_{\mathcal B\mathcal B}$ \\ \hline
                     &\mbox{
                    \leavevmode
                    \epsfxsize=3cm
                    \epsffile[190 630 410 720]{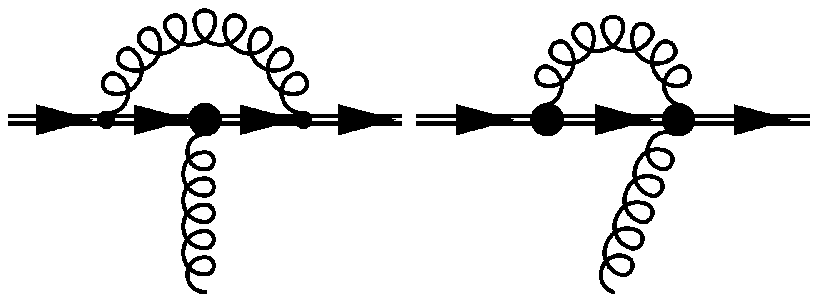}}
&\raisebox{3ex}[1ex]{$  m_{q}\mathcal O^{(2)}_{i}$, $m_{q}^{2}\mathcal
  O^{(1)}_{i}$}&
\raisebox{3ex}[1ex]{$\hat{\gamma}_{\mathcal{O}_{m} \mathcal{O}_{m}}$} \\ \cline{2-4}
\raisebox{6ex}[1ex]{$\vec \mathcal O_{m}$}   &\mbox{
                    \leavevmode
                    \epsfxsize=3cm
                    \epsffile[190 670 410 750]{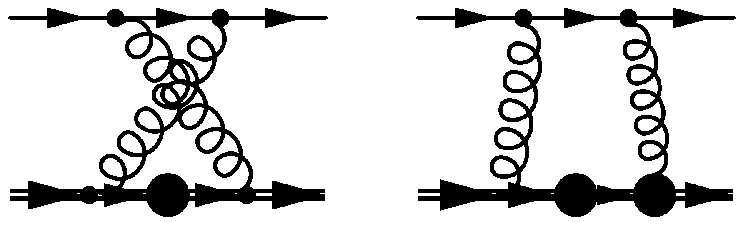}}
&\raisebox{2.5ex}[1ex]{$ m_{q}\mathcal M^{(2)s/o}_{i}\rightarrow\mathcal M^{(3h/l)s/o}_{i\pm}$
(light EOM)}&
  \raisebox{2.5ex}[1ex]{$\hat{\gamma}_{\mathcal{O}_{m} \mathcal M }$} \\ \hline
                   & \mbox{
                    \leavevmode
                    \epsfxsize=1cm
                    \epsffile[260 630 350 740]{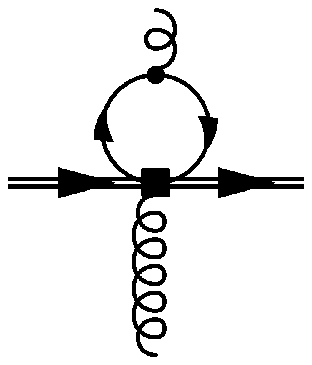}}
&\raisebox{3ex}[1ex]{$ \mathcal O^{(3)}_{i}$, $m_{q}\mathcal O^{(2)}_{i}$,
               $m_{q}^{2}\mathcal O^{(1)}_{i} $}&
  \raisebox{3ex}[1ex]{$ \hat{\gamma}_{\mathcal M \mathcal B}, \hat{\gamma}_{\mathcal M \mathcal{O}_{m}}$}  \\ \cline{2-4}
\raisebox{6ex}[1ex]{$\vec{\mathcal{M}}$}& \mbox{
                    \leavevmode
                    \epsfxsize=1cm
                    \epsffile[250 640 350 740]{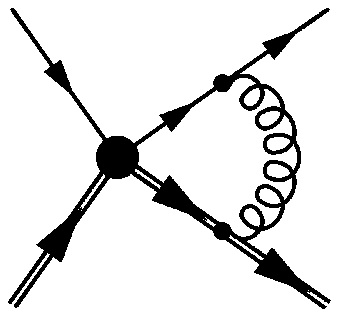}}
&\raisebox{2ex}[1ex]{$ \mathcal M^{(3h/l)s/o}_{i\pm}$}&
   \raisebox{2ex}[1ex]{$\hat{\gamma}_{\mathcal M\mathcal M}$}  \\ \hline
& & $ \mathcal M^{(3h/l)s/o}_{i\pm}$&
   $\hat{\gamma}_{\mathcal T\mathcal M}$ \\ \cline{3-4}
\raisebox{2ex}[1ex]{$\vec{\mathcal{T}}$ }   & \raisebox{1ex}[2ex]{
                    \leavevmode
                    \epsfxsize=1cm
                    \epsffile[250 660 350 740]{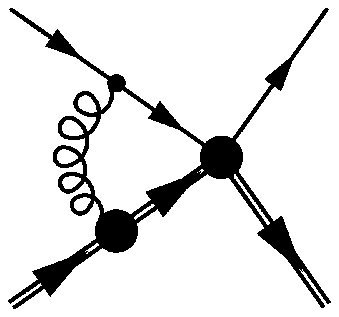}} &
\raisebox{0ex}[0ex]{$ \mathcal M^{(3h/l)s/o}_{i\pm} \rightarrow
                              \mathcal O^{(3)}_{i}$(gluon EOM)}&
 $\hat{\gamma}_{\mathcal T\mathcal B}$ \\ \hline
\end{tabular}
\end{center}
\caption{\label{fig:mix} Mixing properties and operator identities.
The black blob represents an operator of appopriate mass dimension.}
\end{figure}

Since we are working in the background field gauge \cite{BFM,BFM/operators},
it suffices to consider 
1PI Greensfunctions with either two external heavy quark legs
and, depending on the dimension, 
one or two background fields, or two heavy and two light 
fermion fields. Figure \ref{fig:mix} shows  the mixing 
properties of the operator basis
by diagrammatic examples. 
The procedure how to take care of the various operator identities 
consistently follows the lines of \cite{balzereit}.
We present our result for the anomalous dimension matrix corresponding
to the operator basis
$(\vec{\mathcal{B}},\vec{\mathcal{O}}_{m},\vec{\mathcal{M}},\vec{\mathcal{T}})$
as a block matrix:
 \begin{equation}
\label{eq:block}
  \hat\gamma^{(3)} =
      \begin{pmatrix}
      \hat{\gamma}_{\mathcal B\mathcal B} & 0 &\hat{\gamma}_{\mathcal B \mathcal M}&0 \\
    0& \hat{\gamma}_{\mathcal{O}_{m} \mathcal{O}_{m}} &
  \hat{\gamma}_{\mathcal{O}_{m} \mathcal M }&0\\
            \hat{\gamma}_{\mathcal M \mathcal B}&\hat{\gamma}_{\mathcal M \mathcal{O}_{m}}&\hat{\gamma}_{\mathcal M\mathcal M}&0 \\
          \hat{\gamma}_{\mathcal T\mathcal B}&0&\hat{\gamma}_{\mathcal T\mathcal M}&\hat{\gamma}_{\mathcal T\mathcal T} 
      \end{pmatrix}
\end{equation}
The mixing of $\vec \mathcal T$ with $\vec \mathcal B$ is 
a consequence of the gluon EOM removing  
HL--operators in favour of $\mathcal O^{(3)}_{i}$
(see (\ref{eq:gEOMrel})), since to one loop order
there are no penguin diagrams which would require 
$\mathcal O^{(3)}_{i}$ as counterterms.

Furthermore the anomalous dimensions of the H--operators 
(i.e. the entry $\hat{\gamma}_{\mathcal B\mathcal B}$) 
are affected by the gluon EOM since some of the 
$\vec{\mathcal M}$--counterterms may be removed using (\ref{eq:gEOMrel}).
This in turn implies
corrections to the Wilson coefficients of the local operators in 
$\vec{\mathcal B}$.

The operators $\vec \mathcal O_{m}$  
require local counterterms of the form $m_{q} \mathcal M^{(2)s/o}_{i}$ 
which are removed by means of 
(\ref{eq:lightEOM1}) in favour of operators
$\mathcal M^{(3h/l)s/o}_{i\pm}$, thereby explaining
$\hat{\gamma}_{\mathcal{O}_{m} \mathcal M }$. 
The even more complicated situation
that some of the latter obey the gluon EOM fortunately does not occure.

Since the submatrices of the block matrix (\ref{eq:block}) are too large 
we refrain from their presentation. They  
are available from the author on request.

\section{Renormalization group logarithms}
\label{sec:RG-flow}

With the one loop anomalous dimensions and the tree level
matching coefficients we are now in the position
to solve the renormalization group equation for
the Wilson coefficients of the 
effective lagrangian at $\mathcal O(1/m_Q^3)$:
\begin{equation}
\label{eq:RG}
\frac{d}{d\ln \mu}\vec \mathcal C^{(3)} (\mu) + \hat{\gamma}^{(3)\top} \vec
\mathcal C^{(3)}(\mu) = 0
\end{equation}
Here $\vec \mathcal C^{(3)}(\mu)$ denotes collectively the coefficents 
$(\vec C_{\mathcal B},\vec C_{\mathcal O_m},
\vec C_{\mathcal M},\vec C_{\mathcal T})$
of the
physical operators $(\vec{\mathcal{B}},\vec{\mathcal{O}}_{m}
,\vec{\mathcal{M}},\vec{\mathcal{T}})$ in the effective lagrangian
(\ref{eq:eff3}).
Since an analytical diagonalization of $\hat{\gamma}^{(3)}$
seems to be difficult we restrict
ourselves to the calculation of  the first logarithmic
correction $\propto \alpha_s \ln(\mu/m_Q)$ in 
the coefficients $\vec \mathcal C^{(3)}(\mu)$.
The exact solution of (\ref{eq:RG}) reads
\begin{equation}
\label{eq:exact}
\vec \mathcal C^{(3)}(\mu) = \biggl( \frac{\alpha_s(\mu)}{\alpha_s(m_Q)}
                    \biggr)^{\frac{\hat{\gamma}^{(3)\top}}{2\beta^{(0)}}}
                    \vec \mathcal C^{(3)}(m_Q)
\end{equation} 
with the one loop running coupling
\begin{equation}
\frac{\alpha_s(\mu)}{\alpha_s(m_Q)} = 1 - 2\beta^{(0)}
(\frac{\alpha_s(\mu)}{\pi})
\ln(\frac{\mu}{m_Q})
\end{equation}
where $\beta^{(0)} = (33-2n_f)/12$ in the presence of 
$n_f$ light flavors.
Expanding (\ref{eq:exact}) to first order in the strong coupling we get
\begin{equation}
\vec \mathcal C^{(3)}(\mu) = \vec \mathcal C^{(3)}(m_Q) 
           -(\frac{\alpha_{s}(\mu)}{\pi})\ln(\frac{\mu}{m_Q})
             \hat{\gamma}^{(3)\top}\vec \mathcal C^{(3)}(m_Q)
           + \mathcal O((\frac{\alpha_{s}(\mu)}{\pi})^2).
\end{equation}
With our result $\hat{\gamma}^{(3)\top}$ and the tree level 
matching coefficients  
$\vec \mathcal C^{(3)}(m_Q)$ the Wilson coefficients are 
easily calculated. Note that the operators in $\vec \mathcal O_{m}$,
$\vec \mathcal M$ and $\vec \mathcal T$ are not
present at tree level, i.e. the corresponding entries 
in $\vec \mathcal C^{(3)}(m_Q)$ are zero. In figure \ref{fig:WC} the 
non vanishing coefficients are shown.
\renewcommand{\arraystretch}{1}
\begin{figure}[ht]
\begin{center}
\begin{tabular}{|c|c|c|} \hline
Wilson coefficient & tree level & 
coefficient of $(\alpha_{s}(\mu)/\pi)\ln(\mu/m_Q)$\\ \hline
$C^{(3)}_1$ & $2$ & $-25/3\,C_A -32/3\,C_F$ \\ \hline
$C^{(3)}_2$ & $-1$ & $-1/2\,C_A $ \\ \hline 
$C^{(3)}_3$ & $-1$  & $4\,C_A +8/3\,C_F$ \\ \hline
$C^{(3)}_4$  & $1$ & $-17/6\,C_A $ \\ \hline
$C^{(3)}_5$  & $-2$ & $5/3\,C_A+ 8/3\,C_F$ \\ \hline
$C^{(3)}_6$ & $1$ & ${\bf 1/2\,C_{A}}\quad(-\,C_A)$ \\ \hline
$C^{(3)}_7$ & $1$  & $-13/6\,C_A -8/3\,C_F$ \\ \hline
$C^{(3)}_8$ & $1$ & ${\bf 1/2\,C_{A}}\quad(-\,C_A) $ \\ \hline
$C^{(3)}_9$ & $1$ & ${\bf -\,C_A}\quad(-4\,C_A) $ \\ \hline
$C^{(3)}_{10}$ & $-1$  & ${\bf 3/2\,C_A}\quad(9/2\,C_A) $\\ \hline
$C^{(3)}_{11}$ & $-1$  & ${\bf 3/2\,C_A}\quad(9/2\,C_A) $ \\ \hline
$C^{(3)}_{12}$ & $0$  & $1/12\,C_A$  \\ \hline
$C^{(3)}_{13}$ & $0$ & $-1/3\,C_A$ \\ \hline
$C^{(3h)s}_{7+}$& $0$ & $-2\,C_{A}\,C_{F}+1/2\,C_{A}^{2} + 2\,C_{F}^{2}- 1/2$ \\ \hline
$C^{(3h)s}_{7-}$ & $0$ & $-2\,C_{A}\,C_{F}+1/2\,C_{A}^{2} + 2\,C_{F}^{2}- 1/2$ \\ \hline
$C^{(3h)o}_{7+}$ & $0$ & $ 2\,C_{A} - 4\,C_F$ \\ \hline
$C^{(3h)o}_{7-}$ & $0$ & $ \,C_{A} - 4\,C_F$ \\ \hline
$C^{(3l)s}_{1-}$ & $0$ & $- 20/3\,C_A \,C_F + 5/3\,C_A^2 + 20/3\,C_F^2 - 5/3$ \\ \hline
$C^{(3l)s}_{3-}$ & $0$ & $32/3\,C_A \,C_F - 8/3\,C_A^2 - 32/3\,C_F^2 + 8/3$ \\ \hline
$C^{(3l)o}_{1-}$ & $0$ & $5\,C_A - 40/3\,C_F$ \\ \hline
$C^{(3l)o}_{3-}$ & $0$ & $-8\,C_A + 64/3\,C_F$ \\ \hline
$C^{(3l)o}_{6-}$ & $0$ & $ - \,C_A$ \\ \hline
$C^{(3l)o}_{7-}$ & $0$ & $ \,C_A$ \\ \hline
\end{tabular}
\end{center}
\caption{\label{fig:WC} Wilson coefficients of the physical operator basis.}
\end{figure}
The remainig coefficients are $\mathcal O(\alpha_{s}^{2})$ and  
do not contribute in our approximation.

The coefficients of the time ordered products are given by the product 
of the coefficients of their operator components. 

Note, that the coefficients $C^{(3)}_{i}$ for $i = 6,8,9,10,11$
are modified with respect to the value they would get if 
one disregards the HL--operators (in brackets).
This correction is a consequence of 
the gluon EOM which relates certain HL--operators to 
H--operators (see (\ref{eq:gEOMrel})).
In general the same is true for the coefficients $C^{(3)}_{i}$, $i = 2,3,4$
but to one loop order the two operators  $\mathcal M^{(3h)o}_{4\pm}$
which would cause the corrections 
always appear in the combination
$\mathcal M^{(3h)o}_{4+} - \mathcal M^{(3h)o}_{4-} $
which vanishes identically according to (\ref{eq:gEOMrel}) 
or the first relation in 
(\ref{eq:lightEOM2}).
\section{Conclusions}
\label{sec:conclusions}
In this paper we have completed the one loop renormalization
of the HQET lagrangian at $\mathcal O(1/m_{Q}^{3})$. In addition to 
H--operators which are bilinear in the heavy quark field
HL--operators consisting of two heavy and two light quark fields 
are included in the operator basis. We have shown that 
there exist several interdependencies between the 
operators of the complete operator basis which makes its reduction to a 
basis of linearely independent operators a nontrivial task. 
The anomalous dimensions of the H--operators have already been calculated
in \cite{balzereit}. However  in the presence of HL--operators 
the short distance coefficents of the H--operators are modified. 
This is a consequence of the gluon EOM which relates certain HL--operators to
H--operators thereby modifying their anomalous dimensions and 
coefficients. This effect is well known at $\mathcal O(1/m_{Q}^{2})$
where it causes corrections to the coefficient of the Darwin operator
\cite{bauer}.
However, this correction
is weak of $\mathcal O(\alpha_{s}^{2})$ whereas at $\mathcal O(1/m_{Q}^{3})$
the coefficients receive corrections already at $\mathcal O(\alpha_{s})$.

\section*{Acknowledgements}
This work is supported by  
Graduiertenkolleg ``Elementarteilchenphysik an Beschleunigern''.
The author wish to thank T. Mannel for useful and enlightening discussions.


\end{document}